\documentclass[12pt]{article}
\usepackage{graphicx}
\usepackage{epsfig}
\usepackage{mathtext}
\usepackage[T2A]{fontenc}
\usepackage[koi8-r]{inputenc}
\newcommand{\beq}{\begin{equation}}
\newcommand{\eeq}{\end{equation}}
\newcommand{\beqn}{\begin{eqnarray}}
\newcommand{\eeqn}{\end{eqnarray}}
\begin{document} 
 
\title{\textbf{WHAT ARE THE NEUTRINO MASSES. DARK MATTER}} 
\author{V.P. Efrosinin\\
Institute for Nuclear Research\\
Moscow 117312, Russia}

\date{}
\renewcommand {\baselinestretch} {1.3}

\maketitle
\begin{abstract}
The arguments connecting detections of a reason of difficulties of a solution of
a problem of a cold dark matter are adduced.
    
\end{abstract}

Earlier in \cite{efr1} we considered the leptonic vertex of electroweak theory:
\begin{eqnarray}
\label{eq:M1}
L^{CC}=-\frac{g}{2\sqrt{2}}j^{CC}_{\alpha}W^{\alpha}+h.c., 
\end{eqnarray}
where $g$ is the SU(2) gauge coupling constant and CC current is
\begin{eqnarray}
\label{eq:M2}
j^{CC}_{\alpha}=2\sum_{l=e,\mu,\tau}\bar{\nu}_e\gamma_{\alpha}l_L. 
\end{eqnarray}

Thus the reaction was studied                          
\begin{eqnarray}
\label{eq:M3}
\nu_e+n \to e^{-*}+p, 
\end{eqnarray}
\begin{eqnarray}
\label{eq:M4}
e^{-*} \to e^-+\gamma, 
\end{eqnarray}
that is the reaction
\begin{eqnarray}
\label{eq:M5}
\nu_e+n \to e^-+p+\gamma. 
\end{eqnarray}
In \cite{efr1} the parameters of reaction
(\ref{eq:M3}) with an electron off-mass-shall: mass of an exited electron,
angular distribution, cross section and so on were studied.
It should be noted that the process (\ref{eq:M3}) and reaction
\begin{eqnarray}
\label{eq:M6}
\nu_e+n \to e^-+p, 
\end{eqnarray}
where $e^-$ is on the mass shell, are not coherent.

In reaction (\ref{eq:M6}) the leptonic vertex of the Lagrangian
(\ref{eq:M1}),(\ref{eq:M2}) is on a mass surface. And in reaction
(\ref{eq:M3}) this vertex is half outside of a mass surface. There is a
production of an
exited (virtual) electron. The experiment demonstrates, that thus the leptonic
vertex behaves similarly to as well as in reaction
(\ref{eq:M6}).
Anyway it is not accepted in it to enter any form factors. It is difficult
to present that the vertex
(\ref{eq:M1}),(\ref{eq:M2}) acts as a certain semiconductor.
When in one direction the production of the off-shell electron is possile.
And in the other direction there is no production of the neutrino in an exited
state.
Therefore it is impossible to eliminate and capability of production a virtual
neutrino in reverse reaction
\begin{eqnarray}
\label{eq:M7}
e^-+p \to \nu^*_e+n, 
\end{eqnarray}
where $\nu^*_e$ is neutrino off-mass-shell. And the perameters of this reaction
can be like to parameters of reaction
(\ref{eq:M3}), including mass an exited neutrino   
$\nu^*_e$ and cross-section. And the reaction (\ref{eq:M7})
will be incoherent of the reaction
\begin{eqnarray}
\label{eq:M8}
e^-+p \to \nu_e+n, 
\end{eqnarray}
as the direct reactions (\ref{eq:M3}) and (\ref{eq:M6}) are incoherent.
The exited neutrino $\nu^*_e$, as against $\nu_e$, is outside of a mass
surface, and therefore has other nature.

Thus there can be a rather large reduction of cross-section the production
exited neutrino, than as contrasted to by reaction (\ref{eq:M5}).
As the exited electron rather fast descends  from off-mass-shell at production
of a photon. And exited neutrino long live as roaming propagator.

So at production the neutrino with small mass in some reaction is possible also
production a neutrino with large mass in unbroken spectrum.The uncertainty
principle reduces cross section of production a neutrino with large mass,
probably, and is very strong.

The mechanism of operating of the uncertainty principle is not known. Can take
place and absence of a capability the neutrino in general to be in an exited
state.

Further we shall make an estimation about reducing effect of the uncertainty
principle by consideration decay      
\begin{eqnarray}
\label{eq:M9}
K^+ \to \mu^{+*}+\nu_{\mu}, 
\end{eqnarray}
\begin{eqnarray}
\label{eq:M10}
\mu^{+*} \to \mu_{\mu}+\gamma. 
\end{eqnarray}

\begin{figure*}[hb]
\begin{center}
\epsfig{file=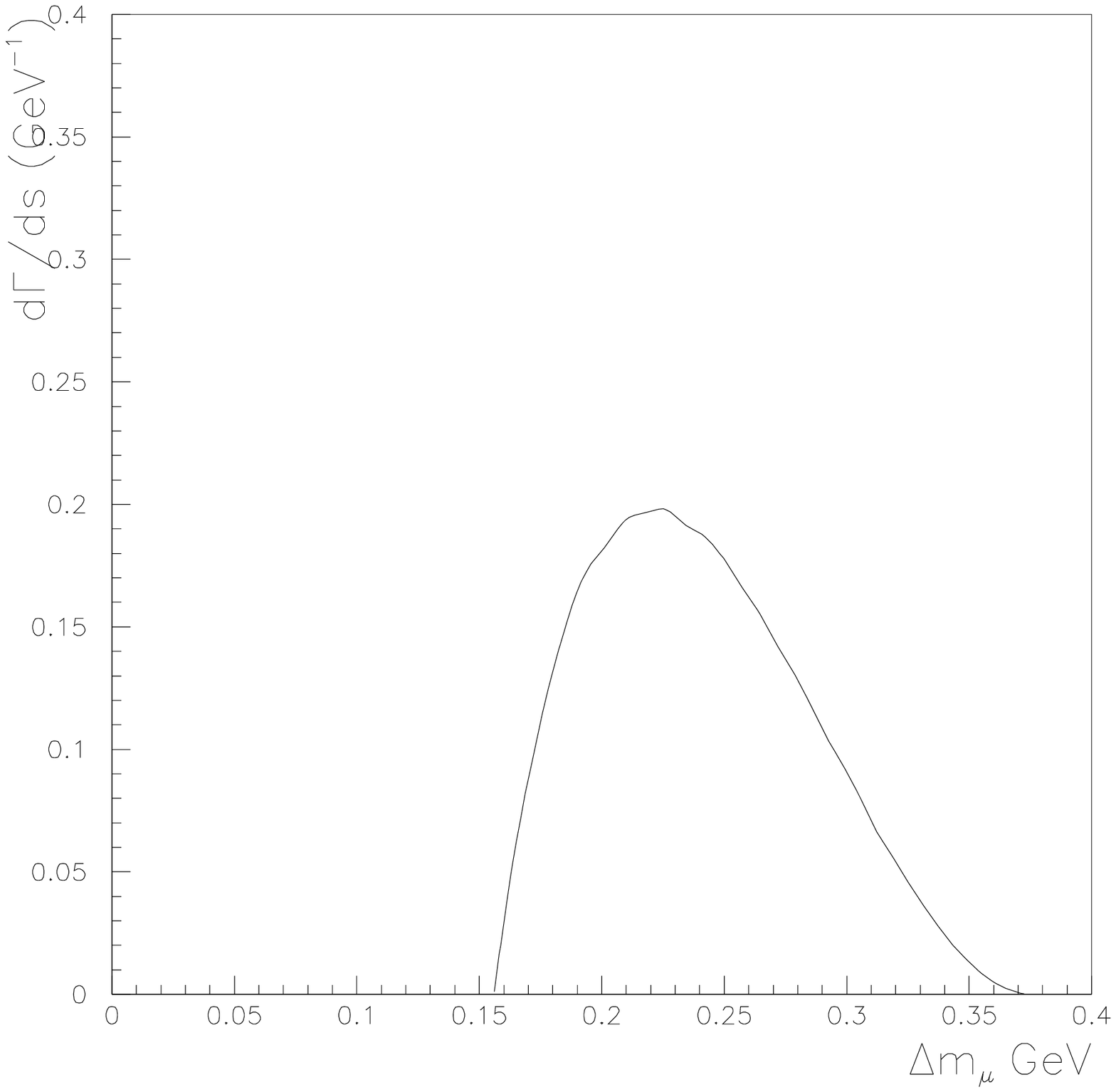,width=14.cm}
\end{center}
\caption{}
\label{fig:fi14}
\end{figure*}

In Fig.\ref{fig:fi14} the distribution of width of decay of
$K^+$ meson from a square of effective mass
$s$ of an exited $\mu^{+*}$ is adduced.
On an abscissa axis the value
$\Delta m_{\mu}=\sqrt{s}-m_{\mu}$ is adduced.
By analogy with the previous arguing the following decay can be watched also
\begin{eqnarray}
\label{eq:M11}
K^+ \to \mu^++\nu_{\mu}^*. 
\end{eqnarray}

From Fig.\ref{fig:fi14} it is visible, that the leading edge of shedule with
mass an exited neutrino
$\sim 0.2$ GeV will penetrate through prohibiting operating of uncertainty
principle. In definite sense it will be a certain mean mass
$\nu^*$ of three satellites for three generations neutrinos. However, all this
has only quality nature, creating reference points for further comprehension.

If the production $\nu^*$ takes place, it is possible, that it will give the
contribution to a cold dark matter.

Within the limits of a nonrelativistic dilute gas the distribution function
receives Boltsman kind. In simplification version
($T\ll m$) the expression for density of number of particles becomes 
\cite{rubak} (further we use our calculations thys monography):
\begin{eqnarray}
\label{eq:M12}
n_i=g_i\Bigl(\frac{m_i T}{2 \pi}\Bigr)^{3/2}\exp{(-m_i/T)}, 
\end{eqnarray}
and density of energy is
\begin{eqnarray}
\label{eq:M13}
\rho_i = m_i n_i.
\end{eqnarray}
Here $g_i$ is number of degree of freedoms, for a neutrino
$g_i$ = 6, $m_i$ - mass of heavy neutrino
$\nu^*$, we take $m_i$
= 0.2 GeV, $T$ is the present temperature.
Supposing that $\rho_i = 0.2 \rho_c$, $\rho_c$ is the critical density,
$\rho_c = 0.53\times10^{-5}\frac{\textup{GeV}}{\textup{cm}^3}$,
we receive from equations (\ref{eq:M12}), (\ref{eq:M13})
$T_{\nu^*} \simeq 2~ \textup{MeV}$. For baryons is received
$T_B \simeq 10$ MeV,
at $g_B=2$, $m_B=1$ GeV.
Let us remark, that the temperatures $\nu^*$ and baryons are proportional to 
their masses.

We do a crude estimate about cross section of production
$\nu^*$. Thus we are suspected that the density of number of particles
a heavy and light neutrino are proportional to their cross sections of
production or width of decais.
Just,
\begin{eqnarray}
\label{eq:M14}
\frac{n_{\nu^*}}{n_{\nu}}=x \times \alpha.
\end{eqnarray}
Here $n_{\nu^*}=0.53 \times 10^{-5}~\textup{cm}^{-3}$,
$n_{\nu}=112~\textup{cm}^{-3}$, $x$ - the reducing factor of the uncertainty
principle,
$\alpha$ - the fine-structure constant. Thus,
$x \simeq 10^{-5}$.

The detection of such small cross sections on experiment is represented unreal.
From here and difficulty with detection of the candidates on a role of a cold
dark matter.
In a case, considered by us, of difficulty adds an unbroken spektrum of heavy
neutrino
$\nu^*$. The problem of small cross sections of production of candidates on the
cold dark matter remains and in discrete spectrum.

In the present article the arguments concerning problem of a cold dark matter
are addused.
More precisely than problem of a development on experiment of new massive
particles of the special nature.
In our article is showed, that for masses of these particles in the field of
hundreds MeV, there is an experimental insuperability of their detection owing
to small cross section of production.

\newpage
\clearpage

\newpage
\clearpage

\begin{center}
Figure captions
\end{center}

Fig.~1. Distribution of width of decay of $K^+$ meson from a square of
effective mass $s$ of an exited $\mu^{+*}$.


\end{document}